\documentclass[aps,prl,reprint,showpacs,superscriptaddress]{revtex4-1}

\usepackage{graphicx}
\usepackage{tikz}
\usetikzlibrary{decorations.pathmorphing}
\usepackage[T1]{fontenc}
\usepackage{mathrsfs}
\usepackage{miller}
\usepackage{amsmath} 
\usepackage{amssymb}
\usepackage{amsfonts}
\usepackage{xspace}
\usepackage{braket}
\usepackage{siunitx}
\usepackage{hyperref} 
\usepackage{float}
\usepackage[caption=false]{subfig}

\usepackage{silence}
\DeclareSIUnit\electrons{e\textsuperscript{-}}
\DeclareSIUnit\neutrons{neutrons}
\DeclareSIUnit\ppm{ppm}
\DeclareSIUnit\ppb{ppb}
\DeclareSIUnit\lines{l}
\DeclareSIUnit{\calorie}{cal}
\newcommand{\oneelec}[1]{\ensuremath{\mathrm{#1}}}

\newcommand{\vect}[1]{\ensuremath{\boldsymbol{#1}}}

\newcommand{\spinstate}[3]{\ensuremath{{}^{#1}\mathrm{#2}_{#3}}}

\newcommand{\Cthir}{\ensuremath{^{13}\mathrm{C}}\xspace}
\newcommand{\Nfif}{\ensuremath{^{15}\mathrm{N}}\xspace}
\newcommand{\Nfour}{\ensuremath{^{14}\mathrm{N}}\xspace}

\newcommand{\NfifPtwoNeutral}{\ensuremath{^{15}\mathrm{N}_{3}\mathrm{V}^{0}}\xspace}

\newcommand{\NfifPtwoBold}{\ensuremath{\mathbf{{}^{15}N_{3}V}}\xspace}

\newcommand{\PtwoCons}{\ensuremath{\mathrm{N}_{3}\mathrm{V}}\xspace}
\newcommand{\PtwoConsNeutral}{\ensuremath{\mathrm{N}_{3}\mathrm{V}^{0}}\xspace}
\newcommand{\PtwoConsMinus}{\ensuremath{\mathrm{N}_{3}\mathrm{V}^{-}}\xspace}
\newcommand{\PtwoConsPlus}{\ensuremath{\mathrm{N}_{3}\mathrm{V}^{+}}\xspace}

\newcommand{\PtwoConsNeutralBold}{\ensuremath{\mathbf{{N}_{3}V^{0}}}\xspace}
 
\newcommand{\NVnb}{\ensuremath{\mathrm{NV}}\xspace} 

\newcommand{\NVminus}{\ensuremath{\NVnb^{-}}\xspace}

\newcommand{\NV}{\NVnb\xspace}

\newcommand{\Ns}{\ensuremath{\mathrm{N_{s}}}\xspace}

\newcommand{\Nsneutral}{\ensuremath{\mathrm{N_{s}}^{0}}\xspace}
\newcommand{\NsPlus}{\ensuremath{\mathrm{N_{s}}^{+}}\xspace}
\newcommand{\NfourNSub}{\ensuremath{^{14}\mathrm{N_{s}}^{0}}\xspace}

\newcommand{\NfifNSub}{\ensuremath{^{15}\mathrm{N}_{\mathrm{s}}^{0}}\xspace}
\newcommand{\NfifNSubPlus}{\ensuremath{^{15}\mathrm{N}_{\mathrm{s}}^{+}}\xspace}

\newcommand{\NsneutralBold}{\ensuremath{\mathbf{N_{s}^{0}}}}
\newcommand{\NfifNSubBold}{\ensuremath{\mathbf{^{15}N_{s}^{0}}}}

\newcommand{\Trigonal}{\ensuremath{\mathrm{C}_{\mathrm{3v}}}\xspace}
\newcommand\scdot{{\mkern 2mu\cdot\mkern 2mu}}

\begin{document}

\title{All-optical hyperpolarization of electron and nuclear spins in diamond}

\author{B.\ L.\ Green}
\email{b.green@warwick.ac.uk}
\author{B.\ G.\ Breeze}
\author{G.\ J.\ Rees}
\author{J.\ V.\ Hanna}
\affiliation{Department of Physics, University of Warwick, Coventry, CV4 7AL,  United Kingdom}
\author{J.-P.\ Chou}
\affiliation{Wigner Research Centre for Physics, Hungarian Academy of Sciences, Budapest, PO. Box 49, 1525, Hungary}
\author{V.\ Iv\'ady}
\affiliation{Wigner Research Centre for Physics, Hungarian Academy of Sciences, Budapest, PO. Box 49, 1525, Hungary}
\affiliation{Department of Physics, Chemistry and Biology, Link\"oping University, SE-581 83 Link\"oping, Sweden}
\author{A.\ Gali}
\email{gali.adam@wigner.mta.hu}
\affiliation{Wigner Research Centre for Physics, Hungarian Academy of Sciences, Budapest, PO. Box 49, 1525, Hungary}
\affiliation{Department of Atomic Physics, Budapest University of Technology and Economics, Budafoki \'ut 8., Budapest, 1111, Hungary}
\author{M.\ E.\ Newton}
\altaffiliation{Corresponding Author}
\email{m.e.newton@warwick.ac.uk}
\affiliation{Department of Physics, University of Warwick, Coventry, CV4 7AL,  United Kingdom}

\begin{abstract}
Low thermal polarization of nuclear spins is a primary sensitivity limitation for nuclear magnetic resonance. Here we demonstrate optically pumped (microwave-free) nuclear spin polarization of \Cthir and \Nfif in \Nfif-doped diamond. \Nfif{} polarization enhancements up to $-2000$ above thermal equilibrium are observed in the paramagnetic system \Nsneutral{}. Nuclear spin polarization is shown to diffuse to bulk \Cthir{} with NMR enhancements of $-200$ at room temperature and $-500$ at \SI{240}{\kelvin}, enabling a route to microwave-free high-sensitivity NMR study of biological samples in ambient conditions. 
\end{abstract}

\maketitle

The enhancement of nuclear polarization is of great importance to nuclear magnetic resonance (NMR) experiments, where the primary sensitivity limit is caused by the small thermal population differences of nuclear spin levels. The development of a general nuclear hyperpolarization technique at arbitrary fields would enable measurement of biomolecules and reaction dynamics that were not accessible by present techniques, while decreasing routine NMR measurement times by orders of magnitude \cite{Griffin2010}. Several approaches to dynamic nuclear polarization (DNP) processes have been demonstrated that enhance nuclear spin polarization; however, the majority are limited to specific fields \cite{Jacques2009,Fischer2013, Wang2013, Falk2015}, low temperatures \cite{Lee2016,Kaplan2015}, specific molecules \cite{Tateishi2014}, or require microwave irradiation of the sample \cite{Tateishi2014, King2015}. Low temperature is particularly problematic for liquid-state biological samples, where freezing leads to loss of spectral resolution \cite{Siemer2012}. Recently, microwave-free optically-pumped DNP (OPDNP) of a diamond containing a high concentration of  the negatively-charged nitrogen vacancy center (\NVminus) has been demonstrated \cite{Scott2016}; however, the electron-nuclear transfer mechanism is not well-understood.

\begin{figure}[H]
	\centering
	\includegraphics[width=0.9\columnwidth]{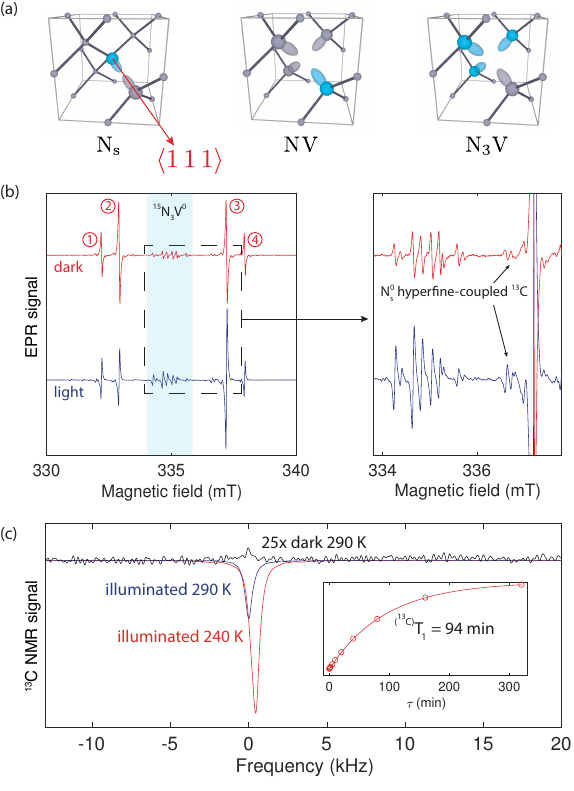}
	\caption{(a) Atomic structures of \Ns{} (left), \NV{} and \PtwoCons{}. In all cases the unpaired electron probability density is localized primarily in the carbon orbitals (gray). (b) EPR spectra collected without (top) and with illumination by \SI{80}{\milli\watt} of light at \SI{532}{\nano\meter} (\SI{2.33}{\electronvolt}) with the sample at \SI{85}{\kelvin} and the external magnetic field $B\||\hkl<111>$. The two visible systems are \NfifNSub (nitrogen hyperfine transitions numbered) and \NfifPtwoNeutral: inversion of the lines under illumination indicates electron spin polarization, and the change in relative intensity of different lines is due to nuclear spin polarization. Panel highlights nuclear polarization of \NfifPtwoNeutral{} and \Cthir{} coupled to \NfifNSub{}. (c) Single shot \Cthir NMR spectra at \SI{7.04}{\tesla}. Illuminated spectra were collected following illumination at \SI{520}{\nano\meter} (\SI{2.38}{\electronvolt}); the dark spectrum was collected after \SI{86}{\hour} at field. Inset: room temperature bulk \Cthir{} polarization build-up, collected after saturating \Cthir{} with a train of $\pi/2$ pulses and illumination for a time $\tau$.}
	\label{fig:spectra_and_structures} 
\end{figure}

In this Letter we demonstrate the electronic spin polarization of two $S=1/2$ paramagnetic nitrogen centers, \Nsneutral{} (substitutional nitrogen [Fig.~\ref{fig:spectra_and_structures}(a)]) and \PtwoConsNeutral{} (vacancy with three nearest-neighbor N [Fig.~\ref{fig:spectra_and_structures}(c)]), in a \Nfif-doped synthetic diamond with an \NVminus concentration \num{<E3} of \Nsneutral{}. Upon illumination, \Cthir{} and \Nfif{} nuclei proximal to the defect centers are spin polarized, with \Nfif{} polarization enhancement of $>2000$ over thermal equilibrium observed. Nuclear spin polarization is shown to diffuse to the bulk \Cthir, leading to microwave-free OPDNP enhancements of $-200$ at room temperature and $-500$ at \SI{240}{\kelvin}. We propose a possible spin polarization mechanism supported by \emph{ab initio} calculations. 

\textit{Sample} -- The sample was grown by the high temperature high temperature (HPHT) method described in \cite{Green2015}, with approximate concentrations of \SI{80}{\ppm} \NfifNSub and

\noindent \SI{5}{\ppm} \NfourNSub. The sample was treated with high energy (\SI{4.5}{\mega\electronvolt}) electron irradiation and HPHT annealing at \SI{1900}{\celsius} to produce \SI{1.6}{\ppm}~\NfifPtwoNeutral, \SI{20}{\ppm}~\NfifNSub and \SI{40}{\ppm}~{N-N} nearest-neighbor pairs. See Supplemental Material for further detail \footnote{\label{footnote:suppInfo}See Supplemental Material at http://abc for sample details, EPR spectra with differently-oriented magnetic fields, and model details including energy level structures for \Nsneutral{} and \PtwoConsNeutral{}.}.

\begin{figure}
	\centering
	\includegraphics[width=0.9\columnwidth]{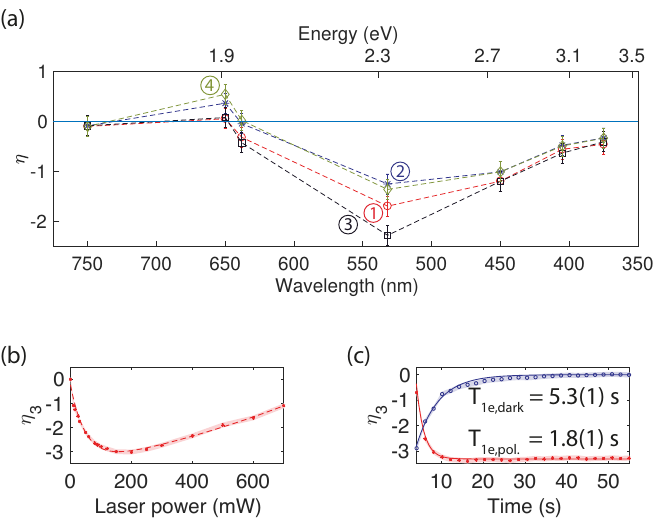}
	\caption{(a) Dependence of EPR enhancement $\eta$ on laser wavelength for each of the \NfifNSub{} hyperfines at \SI{85}{\kelvin} (labeled as Fig.~\ref{fig:spectra_and_structures}(b)). Measurements taken at \SI{80}{\milli\watt} optical power at the sample. (b) EPR enhancement as a function of power at \SI{520}{\nm} and \SI{50}{\kelvin}. (c) Build-up and decay of electron polarization at \SI{50}{\kelvin} when illumination is switched on and off, respectively.}
	\label{fig:lightDependence}
\end{figure}

\textit{Results} -- \Nsneutral and \PtwoConsNeutral centers in diamond each possess a \hkl<111> \Trigonal symmetry axis [Fig.~\ref{fig:spectra_and_structures}(a)], and thus have four symmetry-related orientations. Both centers are $S=1/2$ in the ground state (GS). The use of \Nfif{} ($I=1/2$) during synthesis greatly simplifies the electron paramagnetic resonance (EPR) spectra [Fig.~\ref{fig:spectra_and_structures}(b)] compared to \Nfour{} ($I=1$) due to the lack of nuclear quadrupole interactions \cite{VanWyk1993,Green2017}.

At temperatures below approximately \SI{120}{\kelvin}, in-situ optical illumination results in electron spin polarization of both paramagnetic centers in field-parallel and non-field-parallel orientations [Fig.~\ref{fig:spectra_and_structures}(b)]. The constituent \Nfif nuclei are spin polarized, as are proximal \Cthir (\SI{1.1}{\percent} abundance). The spin polarization mechanism is orientation-dependent \cite{Note1}, and most efficient with $B\|\hkl<111>$ (symmetry axis of one orientation). 

EPR enhancements $\eta = (I_{\mathrm{light}}-I_{\mathrm{dark}})/I_{\mathrm{dark}}$ up to a factor of $\eta=-3$ were measured using \SI{150}{\milli\watt} at \SI{532}{\nano\meter} (\SI{2.33}{\electronvolt}) and a sample temperature of \SI{50}{\kelvin}. The polarization excitation mechanism is highly broadband, with electron and nuclear enhancements measured for \SIrange{750}{375}{\nano\meter} (\SIrange{1.65}{3.31}{\electronvolt}) [Fig.~\ref{fig:lightDependence}(a)]. As the optical power is increased, the polarization saturates before decreasing [Fig~\ref{fig:lightDependence}(b)]: it is postulated that this decrease can be accounted for primarily by a mixture of sample heating and photoionization of \Nsneutral{}.

\begin{figure}
	\centering
	\includegraphics[width=0.9\columnwidth]{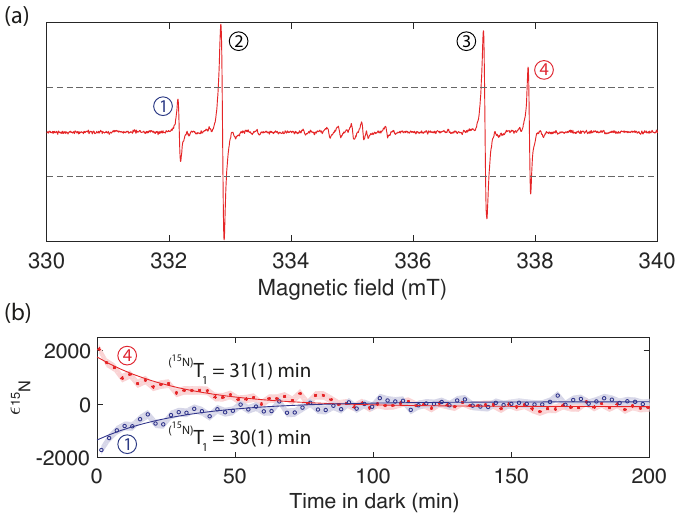}
	\caption{(a) EPR spectrum taken approximately \SI{30}{\second} after illumination is switched off. Field-parallel hyperfine transitions of \NfifNSub{} 1 \& 4 correspond to $\ket{m_{I}}=+1/2$ and $\ket{m_{I}}=-1/2$, respectively: intensity difference is due to \Nfif{} nuclear polarization. Dotted line indicates equilibrium intensity of transitions 1 \& 4. (b) Nuclear polarization of field-parallel \NfifNSub{} hyperfines (1 \& 4) as a function of time: equilibrium is reached with a characteristic lifetime of \SI{31+-1}{\minute} at \SI{50}{\kelvin}. A nuclear polarization of $\epsilon_\mathrm{^{15}N} \approx -2000$ over thermal equilibrium is observed. Hyperfines 2 \& 3 equilibrate with a lifetime of \SI{42+-3}{\minute}. The data have been corrected for a slow charge transfer process (see \cite{Note1}).}
	\label{fig:long_nuclear_lifetimes}
\end{figure}

\Nfif{} nuclear polarization persists after optical excitation is removed, and is strongest in the field-parallel orientation where $m_{S}$, $m_{I}$ are eigenstates of the \Nsneutral{} spin system [Fig~\ref{fig:long_nuclear_lifetimes}]. The difference in relaxation timescales for the electron and nuclei allows the \Nfif spin to be indirectly read-out using the electron. Immediately following the removal of illumination the ratio of observed nuclear polarization to thermal equilibrium, $\epsilon_{\Nfif{}}$, was measured as $-2000$, corresponding to $\approx1/3$ of electron thermal polarization: sequential measurement of the \Nsneutral{} spectrum reveals a nuclear lifetime $^{\Nfif{}}T_{1}=\SI{30+-1}{\minute}$.

Single-shot \Cthir NMR measurements collected with the sample under in-situ optical illumination at \SI{520}{\nano\meter} (\SI{2.38}{\electronvolt}) indicate that the nuclear spin polarization extends beyond the local nuclei and into the bulk [Fig.\ref{fig:spectra_and_structures}(c)]. The characteristic time for this process is $\SI{94}{\minute}$, too slow for an electronic process, and hence is proposed to be mediated by nuclear spin diffusion from the polarized shell around the paramagnetic centers. Bulk OPDNP enhancements of $\epsilon_\mathrm{^{13}C}=\num{-200}$ were measured at room temperature, and $\epsilon_\mathrm{^{13}C}>\num{-500}$ at \SI{240}{\kelvin}, leading to experimental speed-up factors of $40,000$ and $250,000$, respectively. An additional factor of 2 is gained by the reduction in spin-lattice relaxation $^{\mathrm{(^{13}C)}}T_{1,\mathrm{dark}} > \SI{8}{\hour}$ to $^{\mathrm{(^{13}C)}}T_{1,\mathrm{light}} > \SI{1.5}{\hour}$.

\textit{Discussion} -- Two distinct processes can be identified in this sample under illumination: the generation of (electron and nuclear) spin polarization; and the transfer of that polarization out to bulk nuclei. Initially we will not consider how the spin polarization is generated, and simply deal with its transfer to bulk nuclei. Our EPR measurements demonstrate electronic polarization occurring at \PtwoConsNeutral{} and \Nsneutral{} on timescales orders of magnitude faster than the bulk nuclear polarization: we therefore presume that these centers are the source of the polarization. 

\begin{figure}
	\centering
	\includegraphics[width=0.9\columnwidth]{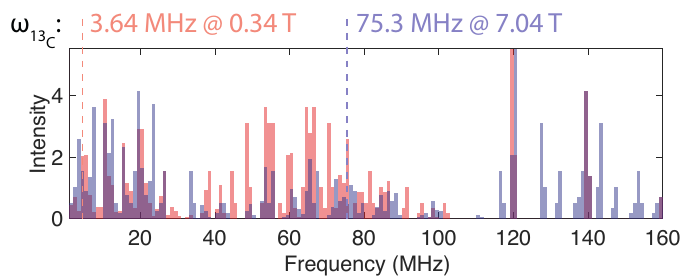}
	\caption{Difference frequencies generated by the ``allowed'' ($\Delta m_{S} = \pm1$; $\Delta m_{I} = 0$) electron transitions of a \NfifNSub{}--\NfifPtwoNeutral{} pair for $B\|\hkl<111>$ at \num{0.34} (red, $\omega_{\Cthir} = \SI{3.64}{\mega\hertz}$) and \SI{7.04}{\tesla} (blue, $\omega_{\Cthir} = \SI{75.3}{\mega\hertz}$) with an isotropic dipolar coupling of \SI{0.5}{\mega\hertz}: stronger couplings will increase the number of frequencies generated and enhance polarization transfer. \Cthir{} hyperfine couplings have been neglected from the model.}
	\label{fig:frequency_comb}
\end{figure}

Several mechanisms exist to transfer polarization from electrons to nuclei, though the typical mechanisms encountered in solids (the solid, cross, and thermal effects \cite{Hu2011,Reynhardt1998}, and Hartmann-Hahn resonance \cite{Knowles2016a}) require microwave driving of the electron spin(s) --- absent in our experiments. We observe nuclear spin polarization at both \num{0.34} and \SI{7.04}{\tesla}, and therefore assume that no resonance coupling of the nuclear and electron spins is required for polarization transfer from electron to nuclei. EPR measurements indicate high levels of nuclear polarization local to the paramagnetic center (within three lattice spacings); however, these nuclei cannot efficiently couple to bulk nuclei due to the local field induced by the electron. Electron spin polarization may be transferred to bulk nuclei via a three-spin electron-electron-nucleus exchange process (i.e. $\ket{+,-,+}\rightarrow\ket{-,+,-}$ in the basis $\ket{m_{S_1},m_{S_2},m_{I}}$), with the condition that the difference of the electron resonance frequencies must equal the nuclear Larmor frequency $|\omega_1 - \omega_2| = |\omega_I|$. At \SI{0.34}{\tesla} and \SI{7.04}{\tesla}, $\omega_{\Cthir} = \num{3.64}$ and \SI{75.3}{\mega\hertz}, respectively. The spin Hamiltonian values for \NfifPtwoNeutral{} and \NfifNSub{} \cite{Green2017} are such that a large number of frequencies between $0$ and \SI{100}{\mega\hertz} are generated at both field strengths [Fig~\ref{fig:frequency_comb}], facilitating polarization transfer out to bulk nuclei. Net bulk polarization will proceed by resonant spin diffusion. This process is sensitive only to the spin Hamiltonian parameters of the interacting defects, and provides a generic route for polarization transfer within dielectric solids.

The above model is sensitive to both the spatial proximity of paramagnetic centers, and also to the spin Hamiltonian parameters of the centers (i.e. the `type' of center). Statistical modeling of relative positions at the present concentrations indicates that between \num{5} and \SI{20}{\percent} of defect center pairs have a separation of \SIrange{1.7}{4.7}{\nano\meter} (see \cite{Note1} for an exploration of model sensitivity to defect center orientation and separation, and magnetic field strength), corresponding to dipolar coupling frequencies of \SIrange{0.5}{10}{\mega\hertz}. This distribution of dipolar couplings will yield a population of centers which are difficult to observe in EPR but will generate additional resonance frequencies (and hence $\Delta\omega_S$), increasing the probability of meeting the polarization transfer matching condition $|\omega_1 - \omega_2| = |\omega_I|$. Additionally, the small difference in $g$-values between the two defects means these conditions will be met for a large range (approx.\ \num{0.3} to \SI{>14}{\tesla}) of magnetic field strengths.

We turn our attention now to the initial generation of the polarization itself. There have been several reports of OPDNP in diamond, however we are aware of only two reports (from the same group) that study all-optical diamond DNP \cite{King2010,Scott2016}: in both cases the effect is attributed to polarization transfer from \NVminus{}. The \NVminus{} concentration in the present sample is below EPR detection limits ($\approx \SI{10}{\ppb}$), even when measured under illuminated (spin-polarized) conditions. Optically-pumped measurements of four other samples, both \Nfour{}- and \Nfif{}-doped with a range of \NVminus{} concentrations [see Table~I in \cite{Note1} for details] failed to exhibit any detectable electron spin polarization: thus we do not attribute the present mechanism to \NVminus{} and must instead consider the other defects present. 

The accepted electronic structure of \Nsneutral{} \cite{Note1} places only one level (of $\oneelec{a_1}$ symmetry) in the band gap: thermoconductivity measurements give the ionization threshold at \SI{1.7}{\electronvolt}, whereas photoionization is subject to a substantial Stokes shift and starts at approximately \SIrange{1.9}{2.2}{\electronvolt} \cite{Heremans2009,Isberg2006}. Similarly, the GS of \PtwoConsNeutral{} has only one hole (also $\oneelec{a_1}$ symmetry), with the excited state transition at \SI{3.0}{\electronvolt} \cite{Walker1979}. Additional transitions at \num{2.6} and \SI{3.6}{\electronvolt} are associated with \PtwoConsNeutral{}: density functional theory (DFT) studies of \PtwoConsNeutral{} suggest they arise from an additional hydrogenic-type state (\PtwoConsPlus{} + $e^{-}$), yielding another $\oneelec{a_{1}}$ state and potentially enabling high-spin ($S>1/2$) states \cite{Jones1997}. Nevertheless, we expect the optical threshold for \PtwoCons{} to be greater than \SI{2.6}{\electronvolt}, contrary to the $\approx\SI{1.9}{\electronvolt}$ observed here [Fig.~\ref{fig:lightDependence}(a)]: these limitations preclude the typical internal singlet-triplet intersystem crossing and level anticrossing polarization mechanisms observed in diamond and SiC \cite{Delaney2010,Ivady2015,Falk2015}. Both \Nsneutral{} (including \NfifNSub{} \cite{Felton2009a}) and \PtwoConsNeutral{} have been studied extensively under optical excitation \cite{Davies1978,Felton2008}, and no spin polarization of either system has been reported. The other high-abundance defects in this sample ($\mathrm{N_{2}}$, $\mathrm{N_{4}V}$) have no reported optical transitions below \SI{4}{\electronvolt}; and the optical absorption spectrum of this sample contains only \Nsneutral and \PtwoConsNeutral \cite{Note1}. 

The simultaneous observation of spin polarization in two well-characterized, optically non-spin polarizable defects suggests a common mechanism. The data allow us to place constraints on such a mechanism: we suppose the same mechanism is responsible for polarization at both \num{0.34} and \SI{7.04}{\tesla}, and therefore is relatively insensitive to magnetic field-strength. Additionally, the mechanism must be capable of spin polarizing electrons and nuclei in multiple systems simultaneously. 

Optical illumination at \SI{>1.9}{\electronvolt} is sufficient to ionize \Nsneutral{}, whereby \PtwoConsNeutral{} centers can capture the carriers and become negatively charged, \PtwoConsMinus{} \cite{Green2017}. Optical absorption measurements of this sample  \Nsneutral{} and \PtwoConsNeutral{} concentrations both increase under \SI{2.33}{\electronvolt} illumination, suggesting the reverse charge transfer process. This is supported by our DFT calculations (see \cite{Note1} for method details), which predict the adiabatic acceptor level of \PtwoConsNeutral{} at $\SI{1.85}{\electronvolt}$ below the conduction band minimum (CBM). Under illumination, the sample is therefore in a metastable equilibrium $(\NsPlus{} + \PtwoConsMinus{}) \leftrightarrow (\Nsneutral{} + \PtwoConsNeutral{})$.

\begin{figure}
	\centering
	\includegraphics[width=0.98\columnwidth]{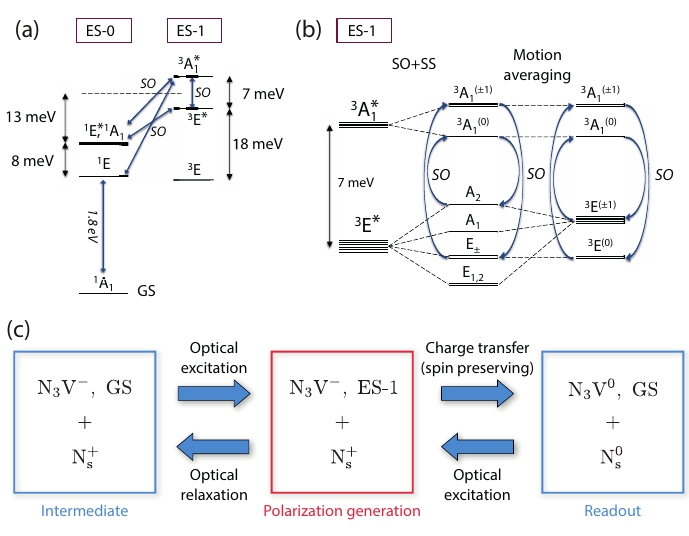}
	\caption{(a) Fine structure of \PtwoConsMinus{} excited states, including the three lowest-energy triplets (ES-1) and singlets (ES-0). The higher energy \spinstate{}{A}{1} and \spinstate{}{E}{} states are marked by $\ast$. Excited states are resonant with the local conduction band minimum. (b) Spin-orbit (SO) coupling effects in the closest pair of $\spinstate{3}{A}{1}^{\ast}$ and $\spinstate{3}{E}{}^{\ast}$ states. Blue arrows indicate transverse spin-orbit coupling. At room temperature, phonon induced spin conserving transitions may average out the spin-orbit splitting of the states driven by axial spin-orbit coupling and electron spin-spin (SS) couplings. (c) Possible model for spin polarization generation. Continuous optical excitation and relaxation causes defect pairs to oscillate between different charge and excitation states. Spin-orbit interactions generate spin polarization in the excited state of \PtwoConsMinus{}; thermal excitation out of this state produces a spin-polarized current which is captured by \NsPlus{}, leading to spin-polarized \Nsneutral{} and \PtwoConsNeutral{}.}
	\label{fig:n3v_minus_detail}
\end{figure}

Further \emph{ab initio} calculations indicate that the CBM states split near the defect due to the perturbation potential of the defect. We find that the excited state of \PtwoConsMinus{} is a bound exciton and includes resonant conduction band states [Fig.~\ref{fig:n3v_minus_detail}(a)]. The calculated radiative lifetime of the singlet \spinstate{1}{E}{} is about three times longer than that of \spinstate{1}{A}{1}, thus these states provide a route for differential decay processes. The $\spinstate{3}{E}{}^{\ast}$ ($\spinstate{3}{A}{1}^{\ast}$) can couple to the \spinstate{1}{A}{1} ($\spinstate{1}{E}{}^{\ast}$) excited state by transverse spin-orbit coupling [Fig.~\ref{fig:n3v_minus_detail}(b)]. The corresponding spin substates of $\spinstate{3}{E}{}^{\ast}$ and $\spinstate{3}{A}{1}^{\ast}$ are also coupled by transverse spin-orbit coupling. 

Upon applying an on-axis (positive) external magnetic field the $\spinstate{3}{A}{1}^{\ast}$ and $\spinstate{3}{E}{}^{\ast}$ states will be slightly $m_{S} = +1$ and $m_{S} = -1$ polarized, respectively, due to the asymmetry of the spin-orbit coupling between the different spin states. The asymmetry, and thus the spin polarization, increases with the magnetic field strength (see \cite{Note1} for the parameters used in the calculation). Due to the transverse spin-orbit coupling and the differential decay from the singlet states, the $\spinstate{3}{A}{1}^{\ast}$ state has a longer lifetime than the $\spinstate{3}{E}{}^{\ast}$ state. As a consequence of a possible thermal ionization of the \PtwoConsMinus{} excited state, the electron spin is left spin-up polarized on \PtwoConsNeutral{} and a spin-polarized carrier is ejected into the conduction band that can be captured by a proximate \NsPlus{} defect, thus spin-polarized \Nsneutral{} will form [Fig.~\ref{fig:n3v_minus_detail}(c)].

\textit{Conclusion} -- Our results show that optical pumping can induce electron and nuclear polarization in two paramagnetic systems in diamond with very low \NVminus{} concentration. NMR measurements with in-situ illumination show that the nuclear polarization diffuses out to the bulk \Cthir{}, leading to OPDNP enhancements of up to \num{-500} at \SI{240}{\kelvin}. The two systems involved, \NfifNSub{} and \NfifPtwoNeutral{}, have only $S=1/2$ states accessible, and hence the standard internal triplet intersystem crossing or level anticrossing mechanisms for solid-state polarization \cite{Ivady2015,Falk2015} cannot be responsible here. Our DFT calculations have indicated the presence of a previously-unidentified high-spin state in the excited state of \PtwoConsMinus{}. Furthermore, it may be possible for this state emit a spin-polarized current, spin-polarizing proximal defects. Electron spin polarization is transferred to bulk nuclei by anisotropic three-spin exchange, with a large set of frequencies generated by the interaction between \NfifNSub{} and \NfifPtwoNeutral{}. Our study implies that engineered synthetic nanodiamonds with concentrations designed to maximize the bulk nuclear polarization would provide a general platform for optical hyperpolarization of a target sample, enabling study of new biological and dynamical systems without the requirement for sample shuttling, low temperature or microwave irradiation.

The authors thank H.\ Fedder, M.\ W.\ Doherty, M.\ W.\ Dale and C.~J.~Wedge for helpful discussions. We acknowledge funding from the Engineering and Physical Sciences Research Council (EP/M013243/1 and EP/J500045/1), the Gemological Institute of America, and the EU Commission (FP7 DIADEMS project No.~611143). We thank De Beers Technologies for provision of samples.

\pagebreak
\widetext
\begin{center}
\textbf{\large All-optical hyperpolarization of electron and nuclear spins in diamond: Supplemental Material}
\end{center}
\setcounter{equation}{0}
\setcounter{figure}{0}
\setcounter{table}{0}
\setcounter{page}{1}
\makeatletter
\renewcommand{\theequation}{S\arabic{equation}}
\renewcommand{\thefigure}{S\arabic{figure}}
\renewcommand{\thetable}{S\arabic{table}}
\renewcommand{\thepage}{S\arabic{page}}
\renewcommand{\bibnumfmt}[1]{[S#1]}
\renewcommand{\citenumfont}[1]{S#1}

\section{Production of \texorpdfstring{\NfifPtwoBold}{15N3V0}}
\begin{figure}[ht]
	\centering
	\subfloat[]{\includegraphics[width=0.2\textwidth]{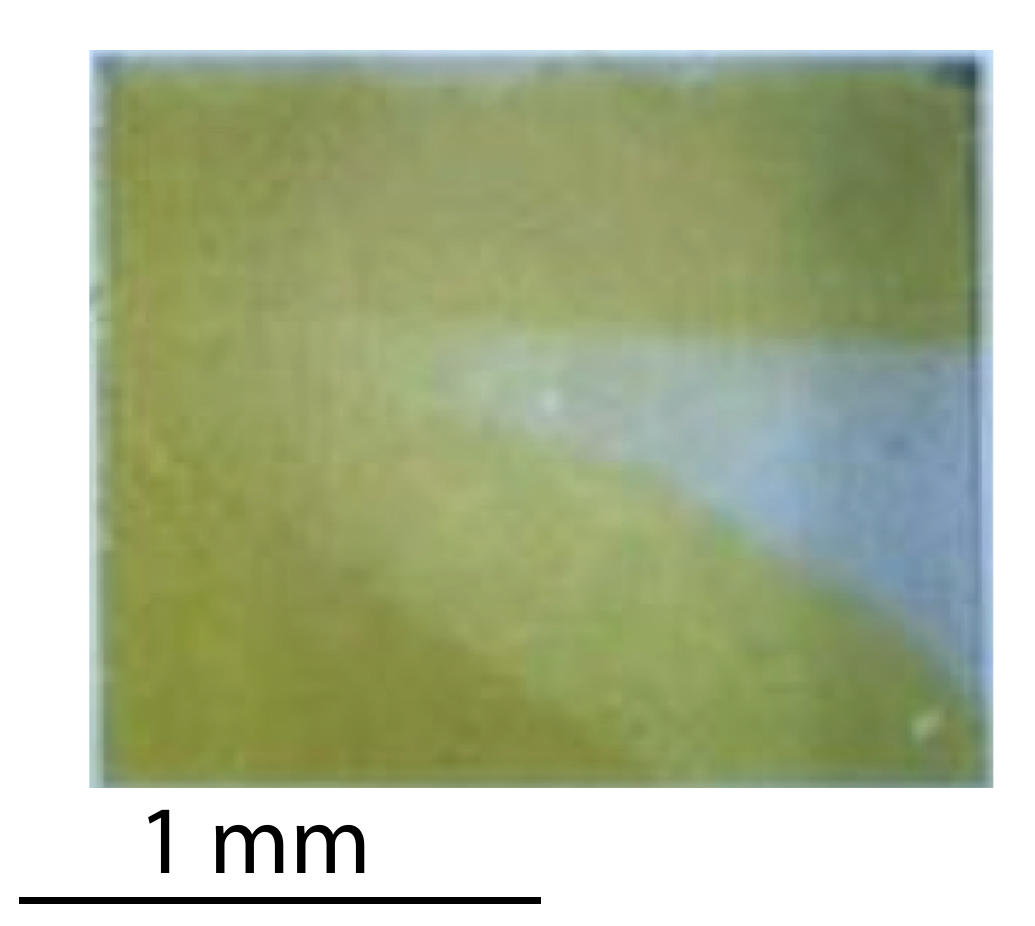}}
	\qquad
	\subfloat[]{\includegraphics[width=0.36\textwidth]{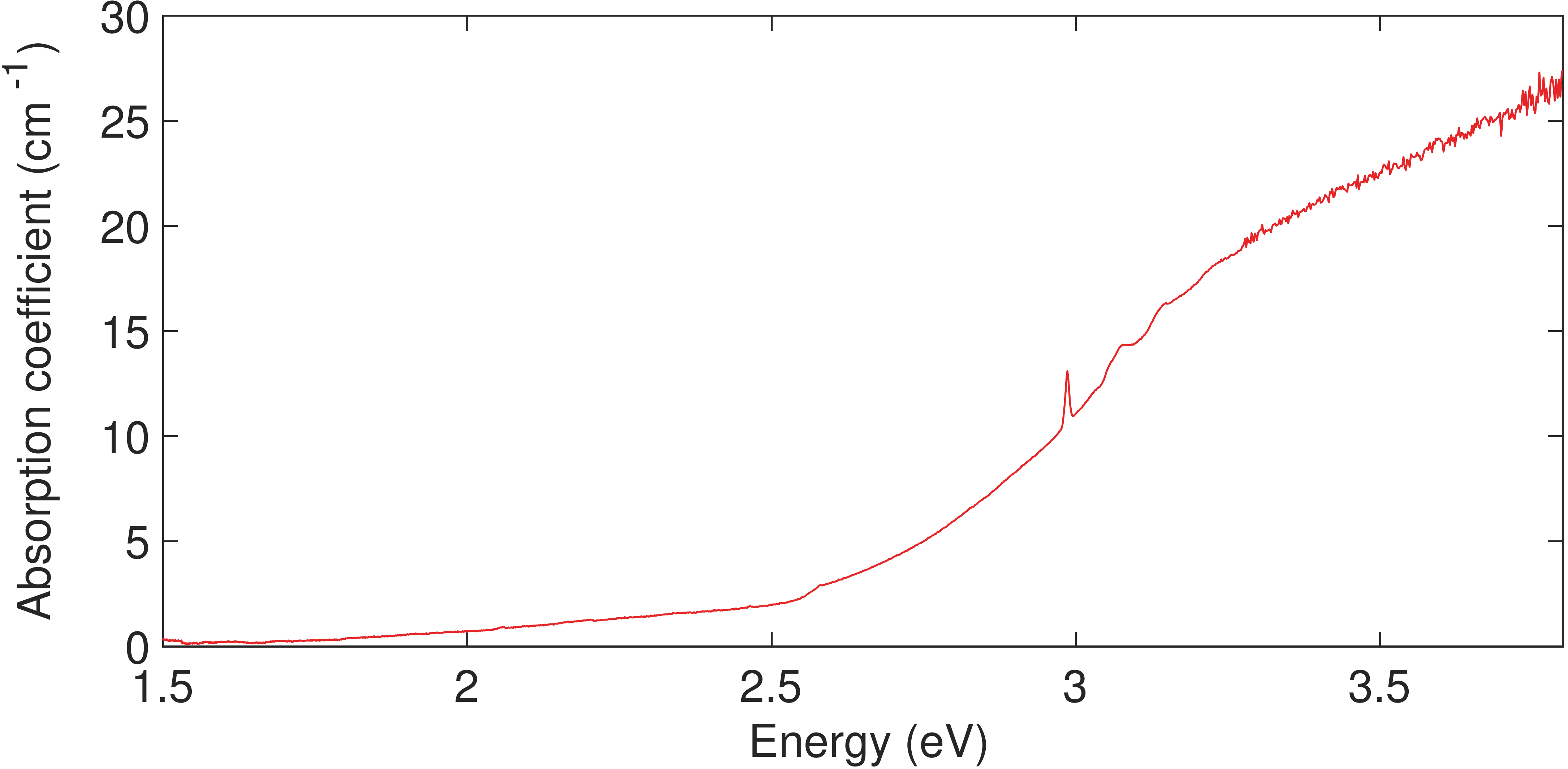}}
	\caption{(a) Photograph of sample used for this study. Nitrogen inhomogeneity is evident by the variation in yellow color saturation in the different growth sectors. Counter-intuitively, the highest concentration of \PtwoCons{} is found in the clear sector: this is because the level of nitrogen aggregation is highest in the high-nitrogen sector, leading directly to a reduction of the yellow color. (b) UV-Vis absorption spectrum of the sample at \SI{80}{\kelvin}.}
	\label{fig:sample}	
\end{figure}

The \Nfif-enriched sample (figure~\ref{fig:sample}) used for EPR and optical studies was grown using the technique described in \cite{S_Green2015}. Post-synthesis, the sample contained mean substitutional nitrogen concentrations of $[\NfifNSub{}] = \SI{80+-2}{\ppm}$ and $[\NfourNSub{}] = \SI{4+-3}{\ppm}$, respectively. The sample was neutron irradiated to a dose of \SI{5E17}{\neutrons\per\centi\meter\squared} and subsequently annealed under a non-oxidizing atmosphere for \SI{15}{\hour} at \SI{1500}{\celsius}, before finally being annealed under high pressure at a nominal temperature of \SI{1900}{\celsius} for \SI{1}{\hour}. This processing regime generated a total concentration of $[\NfifPtwoNeutral] = \SI{1.6+-0.2}{\ppm}$ and substitutional nitrogen concentrations of \SI{20}{\ppm} $[\NfifNSub{}]$ and \SI{5}{\ppm} $[\NfifNSubPlus{}]$, respectively. The sample was polished in order to remove the seed crystal and to provide a flat reference face (within \SI{1}{\degree} of \hkl<110>). Inhomogeneities in the uptake of nitrogen during growth are visible in the sample when viewed under a microscope (figure~\ref{fig:sample}).

\begin{table}[hbt]
	\centering
	\caption{\label{tab:additional_samples}Summary of the samples tested for the presence of electron or nuclear polarization under the same experimental conditions as the primary sample (sample 1).}
	\begin{minipage}[c]{0.8\textwidth}
	\begin{ruledtabular}
	\begin{tabular}{lr@{ :}llllllc}
			Sample	& \multicolumn{2}{c}{Enrichment} & \multicolumn{5}{c}{Defect concentration (ppm)}\\
			 & $\mathrm{^{14}N}$&$\mathrm{^{15}N}$ & \multicolumn{1}{c}{$\mathrm{N_{s}^{0/+}}$} & \multicolumn{1}{c}{$\NVminus{}$} & \multicolumn{1}{c}{\PtwoConsNeutral{}} & \multicolumn{1}{c}{$\mathrm{N_2^{0}}$} & \multicolumn{1}{c}{$\mathrm{N_{4}V^{0}}$} & \multicolumn{1}{c}{NMR measured?} \\
			\hline
			1	& 5&95 	& 25 	& <0.01 	& 1.6 	& 40 	& 15 & Y\\
			2	& 5&95	& 125 	& 			&		&		&    & N\\
			3 	& 5&95 	& 120	& 10		&		&		& 	 & Y\\
			4 	& 15&85	& 38 	& 			& 		& 		&  	 & N\\
			5	& 100&0	& 2		& 0 		& 30	& 		& 	 & Y
			\end{tabular}
	\end{ruledtabular}
	\end{minipage}
\end{table}

Of the five samples measured under the same EPR conditions, three (including the primary sample) were grown simultaneously in the same reaction volume, and hence have the same nitrogen isotopic enrichment (see Table~\ref{tab:additional_samples}): of these, one was measured as-grown, and the other was electron irradiated and annealed to produce \NVminus{} before measurements. Samples 4 \& 5 were HPHT-grown and natural, respectively. 

\section{EPR of \texorpdfstring{\NfifPtwoBold}{15N3V0} and \texorpdfstring{\NfifNSubBold{}}{15Ns0}}
Both \PtwoConsNeutral{} and \Nsneutral{} are common defects in diamond and have been studied extensively in EPR. The parameters used to generate the frequencies for the spin diffusion model, and for fitting of data for extraction of polarization levels are given in table~\ref{tab:spinHamiltonianParameters}.

\begin{table}[H]
	\centering
	\caption{\label{tab:spinHamiltonianParameters}Spin Hamiltonian parameters for the two paramagnetic centers \NfifNSub{} \cite{S_Cox1994} and \NfifPtwoNeutral{} \cite{S_Green2017}. $\theta$ measured from \hkl[110] toward \hkl[001].}
	\begin{ruledtabular}
	\begin{tabular}{lllllllll}
			Center & {$g_{\|}$} & \multicolumn{1}{c}{$g_{\perp}$} & \multicolumn{1}{c}{$\theta_{g}$} & {$A_{1}$} & {$A_{2}$} & {$A_{3}$} & {$\theta_{A}$} \\
			\hline
			\NfifNSub{} & 2.0024 & 2.0024 & 35.26 & -113.838 & -113.838 & -159.73 & 35.26\\
			\NfifPtwoNeutral{} & 2.00241(5) & 2.00326(5) & 35.26 & -10.44(5) & -10.46(5) & -15.85(5) & 157.8(2)\\
			\end{tabular}
	\end{ruledtabular}
\end{table}

\section{Orientation-dependence of spin polarization}
\begin{figure}[h]
	\centering
	\includegraphics[width=0.5\columnwidth]{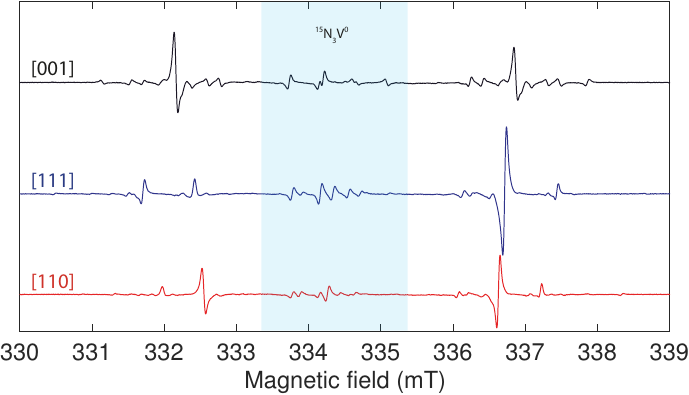}
	\caption{EPR spectra along three high-symmetry directions under illumination from \SI{70}{\milli\watt} of \SI{532}{\nano\meter} light at a sample temperature of \SI{85}{\kelvin}.}
	\label{fig:polarizationRoadmap}	
\end{figure}

The effect of the observed spin polarization depended strongly on the orientation of the external magnetic field $B_{0}$. The effect was strongest with $B_{0}\||\hkl<111>$, where all detectable paramagnetic species exhibited both electronic and nuclear spin polarization; and was weakest for $B_{0}\||\hkl<001>$, where nuclear polarization was detectable on the \Nfif and \Cthir hyperfines of \NfifNSub and the primary hyperfines of \NfifPtwoNeutral (see figure~\ref{fig:polarizationRoadmap}). Contrast with the ``dark'' spectrum from [Fig.~1] of main text.

\section{The model}

\subsection{Electronic structure of \texorpdfstring{\NsneutralBold}{Ns0} \& \texorpdfstring{\PtwoConsNeutralBold{}}{N3V0}}

\begin{figure}[H]
	\centering
	\includegraphics[width=0.4\columnwidth]{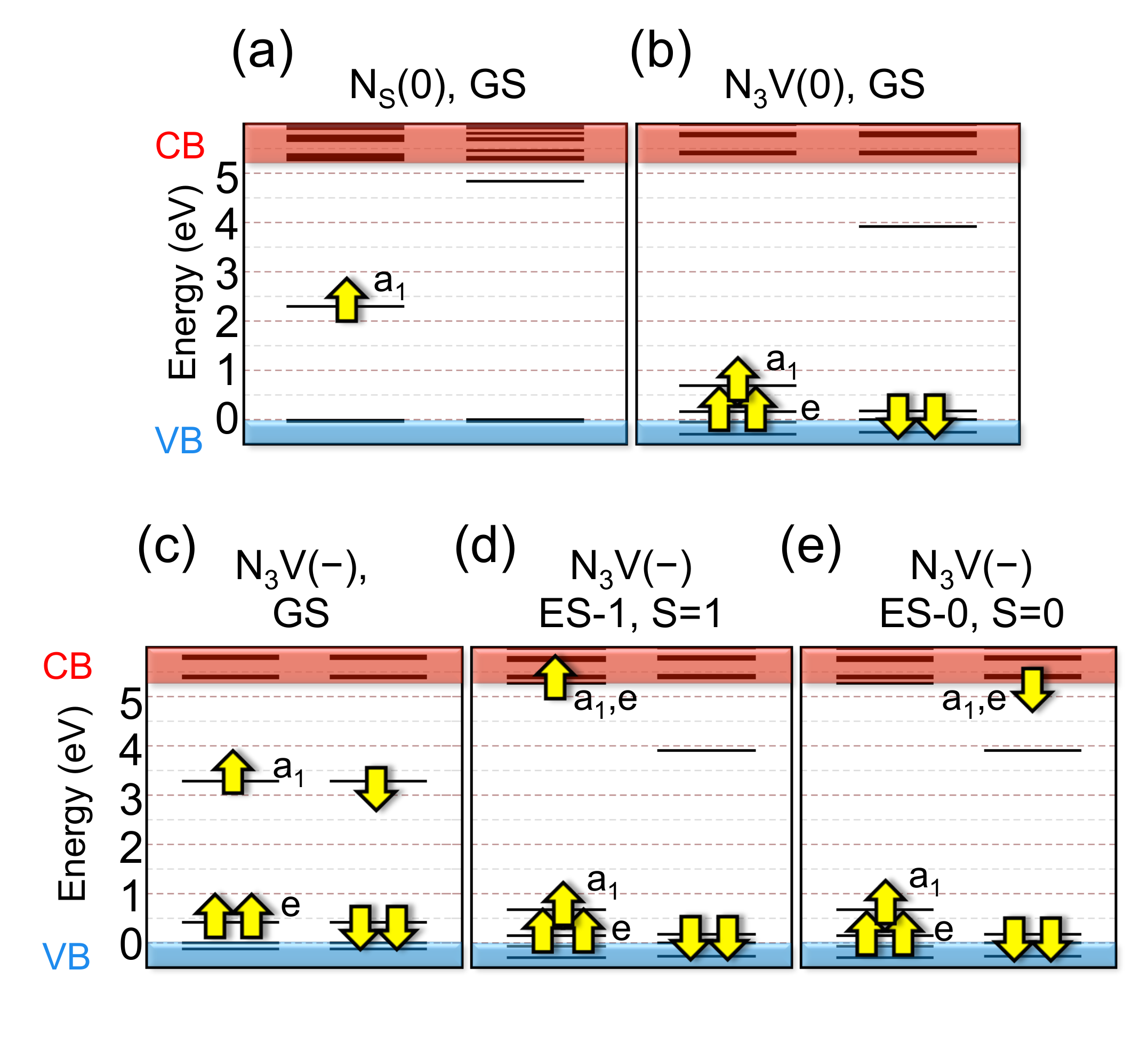}
	\caption{One-electron energy levels for (a) \Nsneutral{} and (b) \PtwoConsNeutral{}. New DFT calculations of the ground and excited states of \PtwoConsMinus{} are given (c--e).}
	\label{fig:energy_levels}
\end{figure}

Due to its role both as one of the most abundant impurities in diamond and its potential as a donor, the electronic structure of \Nsneutral{} has been studied extensively (figure~\ref{fig:energy_levels}). It is well-established from thermoconductivity measurements that the ground state lies approximately \SI{1.7}{\electronvolt} below the band gap \cite{S_Farrer1969a}. Photoconductivity measurements report cut-on thresholds at approximately \SIrange{1.9}{2.2}{\electronvolt} \cite{S_Heremans2009,S_Isberg2006}. There is some suggestion that \Ns{} may also possess an acceptor level, but the transition energy is approximately \SI{4}{\electronvolt} and hence cannot be the intrinsic source of the observed spin polarization \cite{S_Jones2009b,S_Atumi2014}.

The electronic structure of \PtwoCons{} is not definitively known. In the neutral charge state, the ground state and excited state characters (\spinstate{2}{A}{1} and \spinstate{2}{E}{}, respectively) have been experimentally verified via optical characterization of the N3 transition \cite{S_Runciman1965,S_Crowther1967} and EPR of the ground state \cite{S_Davies1978,S_VanWyk1982,S_VanWyk1993}. Some confusion has arisen due to the presence of additional optical transitions (N2, N4) which appear to arise at the same center \cite{S_Davies1978}. The N2 transition was associated with \PtwoConsNeutral{} by  correlation with N3 transition intensity over an order of magnitude in intensity \cite{S_Davies1978}; however, the absorption cross-section is small in comparison to the N3 transition, leading to suggestions that it may arise from a forbidden dipole transition ($\spinstate{}{A}{1} \leftrightarrow \spinstate{}{A}{2}$ in \Trigonal{} symmetry). It is not clear how to generate these states in the ``vacancy-cage'' electronic model (explicitly treating only those orbits directly pointing into the vacancy) typically used to treat vacancy-type defects in diamond \cite{S_Coulson1957,S_Coulson1971}. More recent theoretical analysis suggested the presence of an additional one-electron level outside the vacancy, weakly bound to defect center \cite{S_Jones1997}: the weak N2 transition is then explained by the difference in wavefunction localization between the ground and excited states (see figure~\ref{fig:energy_levels}(b)). Recent experimental results suggest that the N2 and N3 transitions may not correlate in all circumstances \cite{S_Fisher2016}. If correct, the additional one-electron level would enable the generation of high-spin states ($S=3/2$), which are prohibited in the pure vacancy-cage model. In any case, we note that the threshold energy for the first excited state (N2 (\SI{2.6}{\electronvolt}) or N3 (\SI{3.0}{\electronvolt})) is too high to explain the optical dependence of the spin polarization behavior (see main text).

Our study of the electronic structure of \PtwoCons{} places the additional \oneelec{a_{1}} state in the conduction band rather than the gap (see next section for methods). The calculated zero-phonon-line (ZPL) of \spinstate{2}{A}{1}$\leftrightarrow$\spinstate{2}{E}{} transition is \SI{3.07}{\electronvolt}, close to the experimental data at \SI{3.00}{\electronvolt}, associated with N3 center. We found optical transitions, that might be observable in absorption, only at higher energies than N3 ZPL energy. 

\subsection{\emph{Ab initio} calculations}
Theoretical calculations were performed by using density functional theory (DFT). A 512-atom supercell diamond with \SI{370}{\electronvolt} of plane-wave cutoff energy and $\varGamma$-point sampling of the Brillouin zone was used in the calculations. We applied HSE06\cite{S_HSE2003} hybrid density functional which is capable of providing accurate bandgap and defect levels in diamond within \SI{0.1}{\electronvolt}  to experiment\cite{S_deak_accurate_2010}. The electronic transition (zero-phonon line energy) was calculated by the constrained DFT approach\cite{S_CDFT2009}. The imaginary part of the frequency dependent dielectric matrix which represents the absorption spectrum without excitonic effects were calculated without including local field effect\cite{S_gajdos_linear_2006}. The defect's charge transition level, i.e. $(-|0)$, can be determined by the defect formation energies of the neutral and negatively charged states \cite{S_zhang_chemical_1991}. The finite-size effects of supercells associated with electrostatic interactions were corrected using the scheme developed by Freysoldt \textit{et al.}\cite{S_freysoldt2009,S_freysoldt2011}. We calculated the zero-field splitting parameters associated with the electron spin dipole-dipole interaction using our house-build code\cite{S_bodrog_spinspin_2014,S_ivady_pressure_2014}. In the calculation of the hyperfine coupling constants, the core spin polarization within the frozen valence approximation is taken into account\cite{S_yazyev_core_2005,S_szasz_hyperfine_2013}. 

\subsection{Model Hamiltonian approach}
\begin{table}
\caption{\label{table:data} Parameters of the model Hamiltonian given in Eq.~(\ref{eq:H}).}
\begin{ruledtabular}
          \begin{tabular}{lr} 
			 Crystal field splitting ($\Delta$) & \SI{7}{meV} \\[0.2em]
			 Zero-field splitting in \spinstate{3}{A}{1} ($D^{^{3}A_1}$) & \SI{1.115}{GHz} \\[0.2em]
			 Zero-field splitting in \spinstate{3}{E}{} ($D_{\parallel}^{^{3}E}$) & \SI{1.032}{GHz} \\[0.2em]
			 Spin-spin coupling in \spinstate{3}{E}{} ($D_{\perp}^{^{3}E}$) & \SI{0.516}{GHz} \\[0.2em]
			 Axial spin-orbit coupling strength ($\lambda_{\parallel}$) & \SI{10}{GHz} \\[0.2em]
			 Transverse spin-orbit coupling strength ($\lambda_{\perp}$) & \SI{10}{GHz} \\[0.2em]
       \end{tabular}
\end{ruledtabular}
\end{table}
In order to describe the triplet excited states of \PtwoConsMinus{}, we used the following model Hamiltonian
\begin{equation} \label{eq:H}
\hat{H} = \hat{H}_{cf} + \hat{H}_{ss}^{^{3}A_1^{*}} + \hat{H}_{ss}^{^{3}E^{*}} + \hat{H}_{so}^{^{3}E^{*}} + \hat{H}_{so}^{^{3}A_1^{*} \text{--} ^{3}E^{*}} + \hat{H}_{B}
\end{equation}
where 
\begin{equation}
 \hat{H}_{cf}  = \Delta \left| ^{3}A_1^{*} \right\rangle  \left\langle ^{3}E^{*} \right|
\end{equation}
defines the crystal field splitting between the $^{3}A_1^{*}$ and $^{3}E^{*}$ states  whereas
\begin{equation}
\hat{H}_{B} = g_e \mu_B \mathbf{B} \hat{\mathbf{S}}
\end{equation}
accounts for the Zeeman splitting of the spin states, where $g_e$ and $\mu_B$ are the electron $g$-factor and the Bohr magneton, respectively, $\mathbf{B}$ is the external magnetic field, and $\hat{\mathbf{S}}$ is the spin vector operator. Next,
\begin{equation}
\hat{H}_{ss}^{^{3}A_1^{*}} = D^{^{3}A_1^{*}}_{\parallel} \left(\hat{S}_z^2 - \frac{S \left( S + 1\right)}{3} \right) 
\end{equation}
describes the spin-spin interaction in the $^{3}A_1^{*}$ state whereas
\begin{equation}
\hat{H}_{ss}^{^{3}E^{*}} = D^{^{3}E^{*}}_{\parallel} \left(\hat{S}_z^2 - \frac{S \left( S + 1\right)}{3} \right) + D^{^{3}E^{*}}_{\perp}\left( \hat{\sigma}_z \left( \hat{S}_y^2 - \hat{S}_x^2 \right) - \hat{\sigma}_x \left( \hat{S}_y\hat{S}_x + \hat{S}_x\hat{S}_y\right) \right)
\end{equation}
describes the spin-spin interactions in the $^{3}E^{*}$ state, where $\hat{\sigma}_x$ and $\hat{\sigma}_z$ are the Pauli matrices in the basis of orbital states $\left| X \right\rangle $ and $\left| Y \right\rangle $ of $^{3}E^{*}$ \cite{S_Doherty2013}. Finally,
\begin{equation}
\hat{H}_{so}^{^{3}E^{*}} =   \lambda_{\parallel} \hat{\sigma}_y \hat{S}_z
\end{equation}
describes the spin-orbit interaction in the $^{3}E^{*}$ state, where $\hat{\sigma}_y$ is the corresponding Pauli matrix in the basis of orbital states $\left| X \right\rangle $ and $\left| Y \right\rangle $, and
\begin{equation}
\hat{H}_{so}^{^{3}A_1^{*} \text{--} ^{3}E^{*}} = \frac{\sqrt{2}\lambda_{\perp}}{4} \left( \left| E_{+} \right\rangle\left\langle ^{3}A_1^{*} \!\left(-1\right) \right| + \left| A_{2} \right\rangle\left\langle ^{3}A_1^{*} \! \left( 0 \right) \right| + \left| E_{-} \right\rangle\left\langle ^{3}A_1^{*} \! \left(+1\right) \right| + c.c.  \right)
\end{equation}
accounts for the spin-orbit coupling between $^{3}A_1^{*}$ and $^{3}E^{*}$ states, where $ E_{+}$, $ E_{-}$, and $A_2$ and $ ^{3}A_{1}^{*}\! \left( m_{S}\right)$ are the eigenstates of $\hat{H}_{}^{^{3}E^{*}} = \hat{H}_{ss}^{^{3}E^{*}} + \hat{H}_{so}^{^{3}E^{*}} $ and $\hat{H}_{ss}^{^{3}A_1^{*}}$, respectively. Parameters used in the above described model Hamiltonian are given in Table~\ref{table:data}. The spin-spin interaction parameters and the \spinstate{3}{A}{1} -- \spinstate{3}{E}{} gap were obtained by our \emph{ab initio} calculations, while the spin-orbit coupling strengths were chosen to be comparable with the known parameters of \NVminus \cite{S_Batalov2009}.

\begin{figure}[H]
	\centering
	\includegraphics[width=0.4\columnwidth]{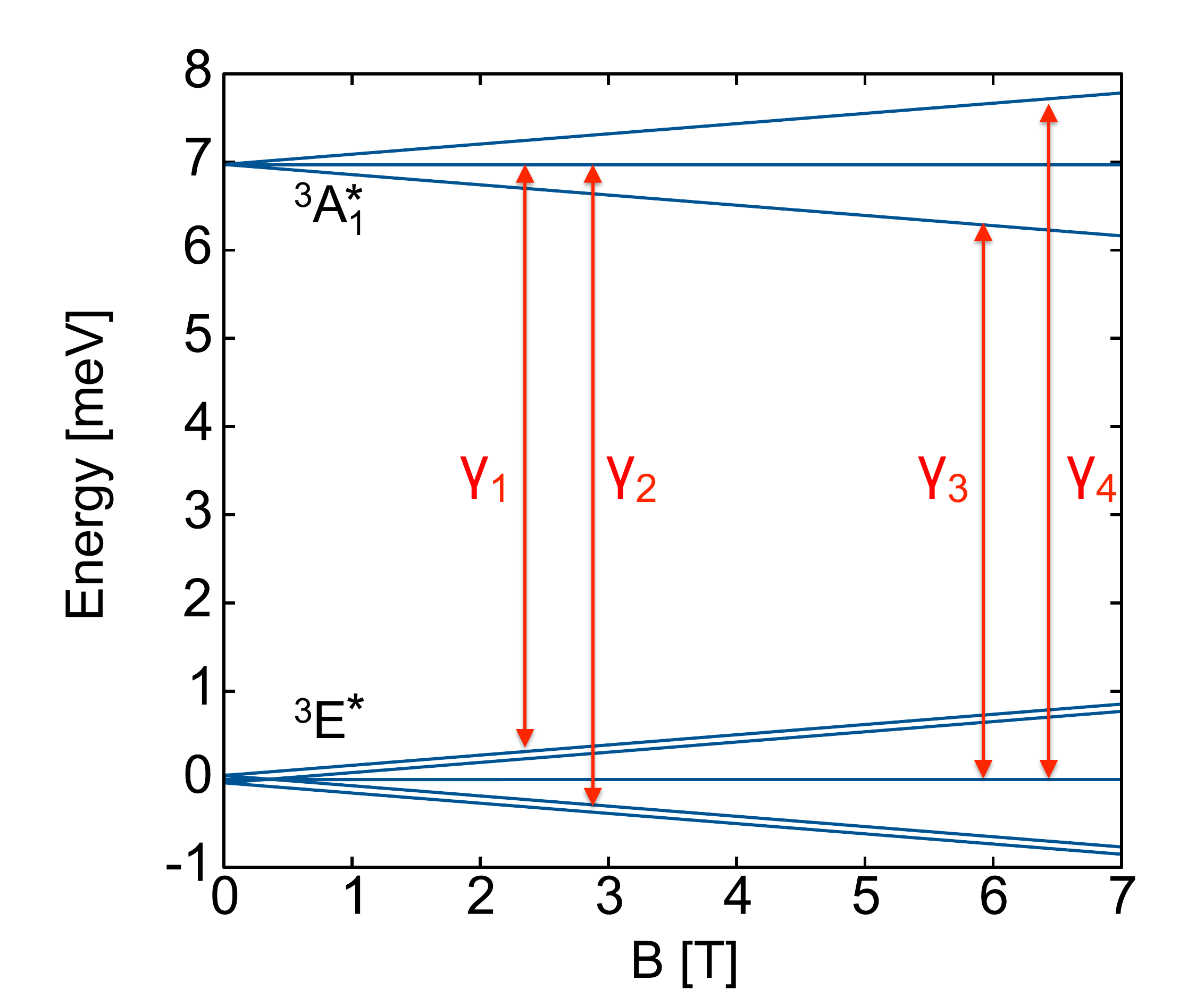}
	\caption{Magnetic field dependence of \spinstate{3}{A}{1}$^{*}$ and \spinstate{3}{E}{}$^{*}$ states. Due to the varying energy gaps of the states in external magnetic field, different spin-orbit mixing (red arrows) can be observed that results in a net spin polarization of the \spinstate{3}{A}{1}$^{*}$ and \spinstate{3}{E}{}$^{*}$ states.}
	\label{fig:fine-field-dep}
\end{figure}

Magnetic field dependence of the eigenstates of our model Hamiltonian $\hat{H} $ is depicted in Fig.~\ref{fig:fine-field-dep}, where the allowed spin-orbit couplings between $^{3}A_{1}^{*}$ and $^{3}E^{*}$ related eigenstates are shown by red arrows. Note that the spin-orbit coupling of the states largely depends on the gap $\Delta$ of $^{3}A_{1}^{*}$ and $^{3}E^{*}$. By applying an external magnetic field, the gaps of different spin-orbit coupled pairs vary, thus the mixing of the states varies too. In order to understand the effect of different coupling strengths caused by an applied magnetic field, we consider the state mixing in $^{3}A_{1}^{*}$ related branch. For positive magnetic fields, the highest energy state is a mixture of  $^{3}A_{1}^{*} \left( +1\right) $ and $ E_{-}$. The larger the magnetic field is, the smaller the coupling strength $\gamma_4$, so the mixing of the states is less pronounced. In other words, the state becomes more $^{3}A_{1}^{*} \left(+1\right) $ like with a defined spin projection of $m_S = +1$. The lowest $^{3}A_{1}^{*}$ related state is a mixture of $^{3}A_{1}^{*} \left( -1\right) $ and $ E_{+}$. For increasing positive magnetic fields the spin-orbit mixing ($\gamma_3$) increases, thus the $^{3}A_{1}^{*} \left( -1\right)$ state loses its $m_S = -1$ character. Using similar arguments, it can be shown that the $^{3}A_{1}^{*} \left( 0\right) $ becomes more  $m_S = +1$ like with increasing magnetic field. As a consequence of all of these effects, the $^{3}A_{1}^{*}$ becomes slightly $m_{S}=+1$ polarized. Inverse effects cause slight $m_S = -1$ polarization of  $^{3}E^{*}$. The above described static polarization process depends linearly on the magnetic field strength as long as $\lambda_{\perp} << \Delta$. 

Differential decay process from $^{3}A_{1}^{*}$ and $^{3}E^{*}$ can result in different lifetime for $m_S = +1 $ and $m_S = -1 $ spin and provide a route for dynamical spin polarization processes to cool down the spins in diamond. Here, we have to consider that the triplet excited states either may be ionized or may scatter to the excited singlet states by transverse spin-orbit interaction mediated by phonons. Since the calculated $^{3}A_{1}^{*}$ energy levels lie higher than the  $^{3}E^{*}$ energy levels, an $m_S = +1 $ spin-polarized current is expected to appear because the spin-up polarized $^{3}A_{1}^{*}$ states can be ionized with higher probability then the spin-down polarized $^{3}E^{*}$ states. If the electron decays from the excited triplet states to the ground state singlet via spin-orbit scattering and radiative decay then the applied optical pumping will repopulate the excited state singlets. According to our calculation the radiative lifetime of the $^{1}E^{*}$ singlets is about twice longer than that of the $^{1}A_{1}^{*}$ singlets. Since the transverse spin-orbit interaction links $^{1}E^{*} \leftrightarrow ^{3}A_{1}^{*}$ and $^{1}A_{1}^{*} \leftrightarrow ^{3}E^{*}$ and $^{1}E^{*}$ singlets have longer lifetime it has a higher probability that the optical pumping leads to $m_S = +1 $ spinpolarization induced by $^{3}A_{1}^{*}$ states. Taken as a whole, we calculate that the electron spin of \PtwoConsNeutral{} will be spin-up polarized, and the spin-up polarized current will lead to a spin-up polarized neighbor \Nsneutral{} defect.  

\subsection{Inter-defect distances for three-spin exchange}
A Monte-Carlo model was constructed in order to estimate the distance between neighboring \Nsneutral{} and \PtwoConsNeutral{} defect centers, using the concentrations of \num{20} and \SI{1.6}{\ppm}, respectively. The ``lattice'' was a cube of 1,024 conventional diamond unit cells per side --- a total of \num{8.6E9} possible atomic sites. Each \Nsneutral{} was placed randomly in the lattice, and for each \PtwoConsNeutral{} the distance to the nearest \Nsneutral{} center was computed: to minimize computation time only the nearest neighbor distance was computed for each center.

\begin{figure}[H]
	\centering
	\includegraphics[width=0.8\textwidth]{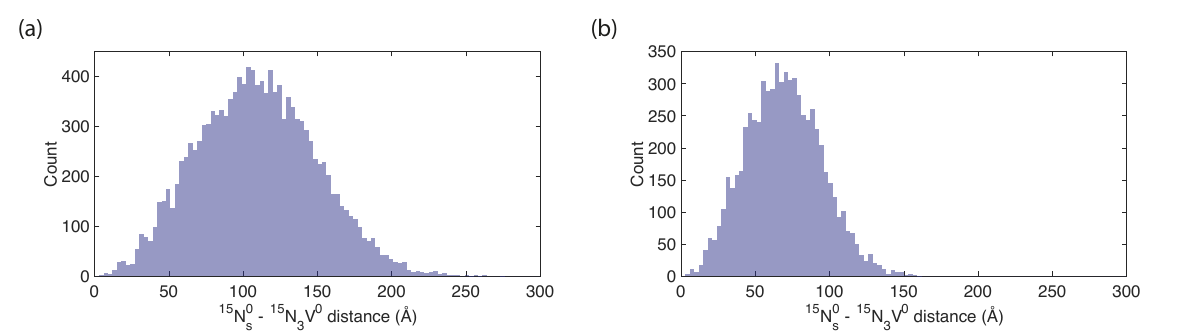}
	\caption{Nearest-neighbor inter-center distances computed for a diamond lattice containing (a) \SI{20}{\ppm} \Nsneutral{} and \SI{1.6}{\ppm} \PtwoConsNeutral{}; (b) \SI{80}{\ppm} \Nsneutral{} and \SI{6.4}{\ppm} \PtwoConsNeutral{}. Distances computed from each \PtwoConsNeutral{} center to the nearest \Nsneutral{} center only.}
	\label{fig:center_distances}
\end{figure}

Two different defect concentration regimes were computed: $\NfifNSub{} = \SI{20}{\ppm}$, $\NfifPtwoNeutral{} = \SI{1.6}{\ppm}$, corresponding to the average distance for the ensemble concentrations measured; and $\NfifNSub{} = \SI{80}{\ppm}$, $\NfifPtwoNeutral{} = \SI{6.4}{\ppm}$, to account for the sectors containing the highest nitrogen density (as estimated from the EPR linewidth). The results are given in figure~\ref{fig:center_distances}.

\subsection{Coupling strengths and orientations}
\begin{figure}[H]
	\centering
	\includegraphics[width=0.8\textwidth]{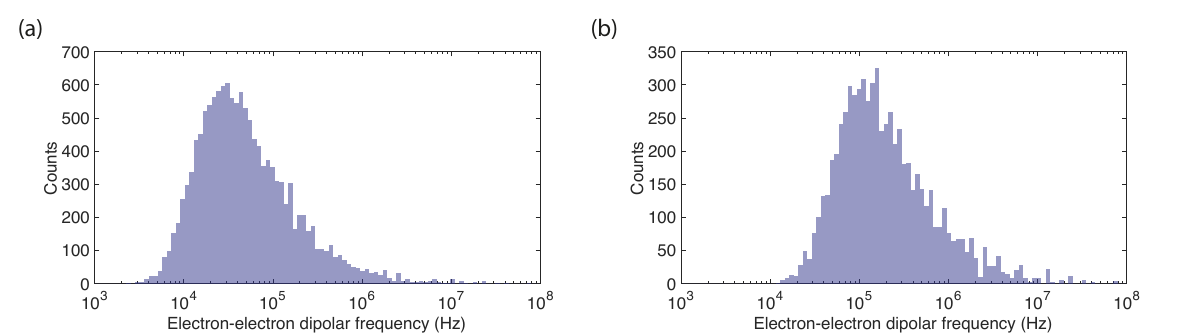}
	\caption{Dipolar coupling frequencies generated by the defect distributions given in figure~\ref{fig:center_distances}.}
	\label{fig:dipolar_coupling_frequencies}
\end{figure}

The spin Hamiltonian relevant to our electron-electron-nuclear three-spin system is
\begin{equation}
	H = \sum_{i}^{2} \sum_{j}^{N} \mu_B \vect{B}^{T} \scdot \vect{g}_{i} \scdot \vect{S}_{i} + \vect{S}_{i}^{T} \scdot \vect{A}_{ij} \scdot \vect{I}_{j} + \vect{S_{1}}^{T} \scdot \vect{J} \scdot \vect{S_{2}}
	\label{eqn:hamiltonian}
\end{equation}
with $i$ and $j$ for electrons and nuclei, respectively. The values of $\vect{A}$ and $\vect{g}$ for \NfifNSub{} and \NfifPtwoNeutral{} are given in Table~\ref{tab:spinHamiltonianParameters}. The value of $\vect{J}$ is dependent on electron-electron separation $r$, with the dipolar contribution (expected to be dominant over exchange at these concentrations) dependent on $1/r^{3}$. Using the inter-defect distributions given in figure~\ref{fig:center_distances}, the corresponding dipolar coupling frequency distribution is generated by
\begin{equation*}
	\omega_{dd} = \frac{\mu_0}{4 \pi h}\frac{g^2 \mu_B^2}{r^3},
\end{equation*}
assuming that the $g$-anisotropy is small \cite{S_Schweiger2001}. Both spatial distributions generate a range of coupling frequencies, with approximately \SI{5}{\percent} of defect pairs in the range \SIrange{0.5}{10}{\mega\hertz} (\SIrange{4.7}{1.73}{\nano\meter} separation) for the low density distribution, and \SI{20}{\percent} for the high density distribution (see figure~\ref{fig:dipolar_coupling_frequencies}). Approximately \SI{1}{\percent} (\SI{3}{\percent}) of pairs in low (high) density regions have a separation of \SI{2.4}{\nano\meter} or less, where the exchange interaction becomes significant compared to dipolar couplings \cite{S_Kortan2016,S_Shulman1967}.

\begin{figure}[H]
	\centering
	\includegraphics[width=0.8\textwidth]{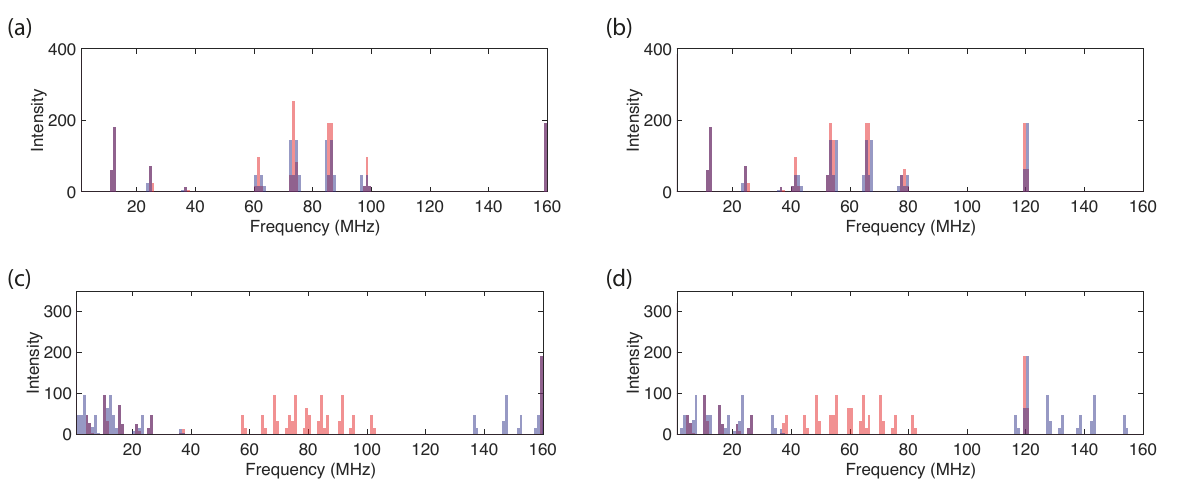}
	\caption{Frequencies generated by solving equation~\ref{eqn:hamiltonian} for different relative orientations of \NfifNSub{}-\NfifPtwoNeutral{} pairs with an isotropic $J = \SI{0.5}{\mega\hertz}$. The defect centers are either parallel ($\|$) to the external field $B_0$, or at an angle $\theta_d = \SI{109.47}{\degree}$ to it (for \hkl[111] and \hkl[-1-11] orientations of each center, respectively). The basis for the following is (\NfifNSub{},\NfifPtwoNeutral{}). {For (a-d), respectively, the orientations are: ($\|$, $\|$); ($\theta_d$, $\|$); ($\|$, $\theta_d$); ($\theta_d$, $\theta_d$).} Figure~4(b) of main text generated by weighting the frequencies generated at each oriented pair by their relative occurrence. All spin Hamiltonian calculations performed using EasySpin \cite{S_Stoll2006}.}
	\label{fig:orientations_and_frequencies}
\end{figure}

\begin{figure}[H]
	\centering
	\includegraphics[width=0.8\textwidth]{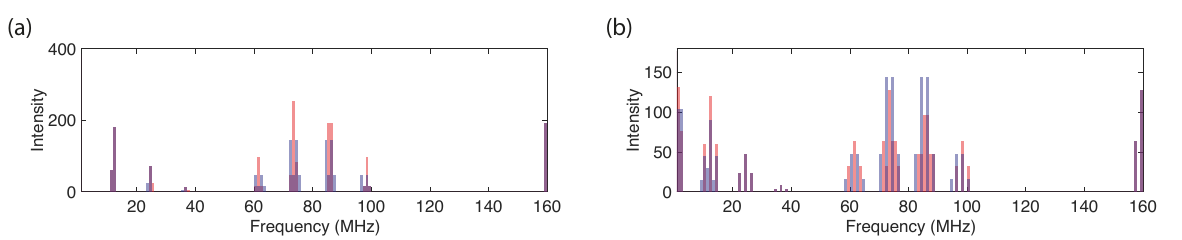}
	\caption{Effect of increasing interaction strength $J$. (a) As figure~\ref{fig:orientations_and_frequencies}(a) ($J=\SI{0.5}{\mega\hertz}$). (b) As (a) with $J=\SI{2.0}{\mega\hertz}$.}
	\label{fig:increasing_dipolar_coupling}
\end{figure}

The spin Hamiltonian parameters for each center are anisotropic (as required by our model \cite{S_Jeschke1997}), and hence the frequencies generated depends on the relative orientations of the defect pair in question. As each center has \Trigonal{} symmetry and we consider only $B_0\|\hkl<111>$, there are only four classes of orientation: both centers parallel to $B_0$; one of the pair parallel to $B_0$; and neither center parallel to $B_0$ --- the frequency contribution of each case is given in figure~\ref{fig:orientations_and_frequencies}. A distribution of dipolar interaction strengths will increase the generated frequencies and hence increase the likelihood that the polarization transfer matching condition (see main text) is satisfied: the effect of changing $J$ is illustrated in figure~\ref{fig:increasing_dipolar_coupling}.

\begin{figure}[H]
	\centering
	\includegraphics[width=0.8\textwidth]{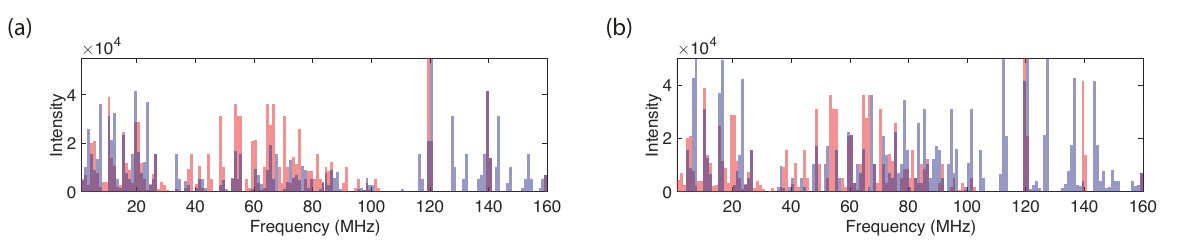}
	\caption{Field-insensitivity: frequencies generated by all four orientations of a \NfifNSub{}-\NfifPtwoNeutral{} pair at (a) \SI{0.34}{\tesla} (red) and \SI{7.03}{\tesla} and (b) \SI{0.34}{\tesla} and \SI{14.0}{\tesla}. An isotropic coupling of $J=\SI{0.5}{\mega\hertz}$ was used in all cases.}
	\label{fig:different_fields}
\end{figure}

\subsection{Polarization efficiency}
The efficiency of the polarization mechanism is difficult to estimate: in our measurements, \SI{40}{\percent} polarization of \SI{5}{\percent} population is indistinguishable from \SI{10}{\percent} polarization of \SI{20}{\percent} population. The sample under study is highly inhomogeneous, with at least three optically distinguishable nitrogen concentrations, and two distinct concentrations visible in EPR spectra (determined by multiple simultaneous linewidths). If the polarization mechanism is dependent on interaction between \Nsneutral{} and \PtwoCons{} then we expect it to occur in only the higher nitrogen sectors (upper limit \SI{40}{\percent} of the sample). At room temperature no electron polarization is visible in the EPR spectra, and the upper limit on \Cthir polarization is therefore given by the ratio of the Boltzmann polarizations $\propto \mu_{e}/\mu_{\mathrm{^{13}C}} \approx \num{2600}$: enhancements of $-200$ correspond to an effective homogeneous efficiency of approximately \SI{8}{\percent}. 

\section{NMR measurements}
The static \Cthir solid state NMR measurements were completed at \SI{7.04}{\tesla} using a Bruker Avance III HD spectrometer. A \SI{5}{\milli\meter} low temperature static probe was used to produce an \SI{80}{\kilo\hertz} $\pi/2$ pulse, which was calibrated on $\mathrm{CH_{3}OH(l)}$. The diamond was mounted into a \SI{3.2}{\milli\meter} $\mathrm{ZrO_2}$ rotor with the \hkl<111> axis parallel to $B_0$. The sample was held in place using the optical fiber fixed into the cap position.

\section{Charge transfer correction}
\begin{figure}[H]
	\centering
	\includegraphics[width=0.6\textwidth]{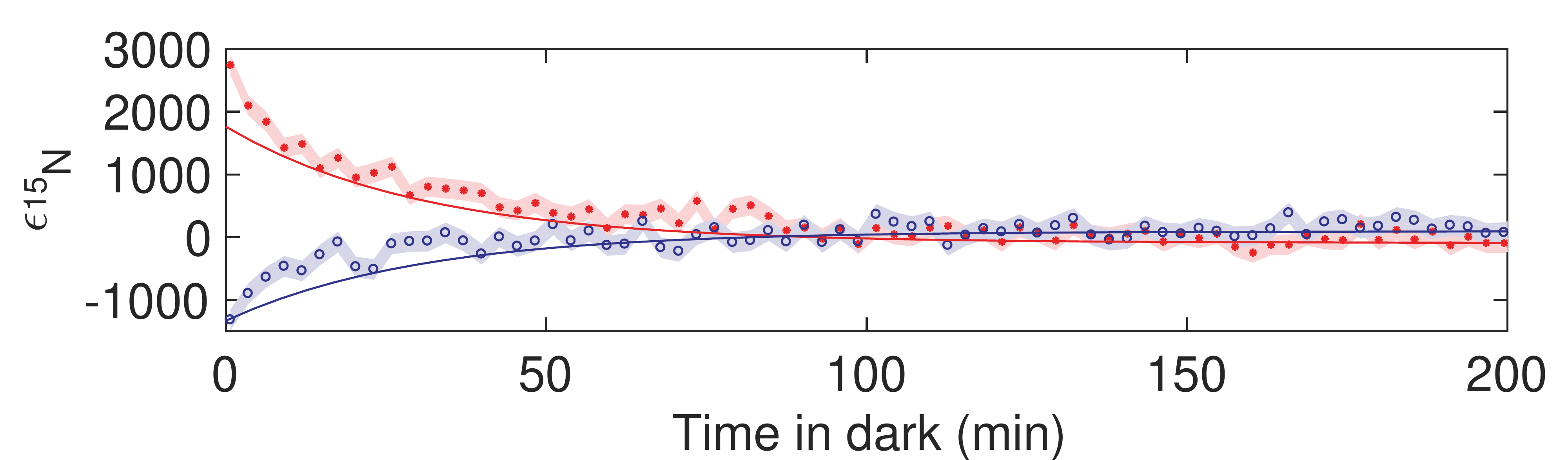}
	\caption{Identical data to figure~3 of main text, but without correction for loss of population during the measurement. On approximately the same timescale as the measurement, the \Nsneutral{} population is exponentially decreasing. The exponential fits from the main text to the (processed) data have been plotted to give a reference for the magnitude of the change.}
	\label{fig:charge_transfer_correction}
\end{figure}

Figure~3 of the main text refers to a correction made for a slow charge transfer process. As detailed in the main text, the light excitation drives charge transfer between \Ns{} and \PtwoCons{} (and potentially other defects). Once the light is switched off, the populations of these centers do not revert immediately back to their pre-light state: there is a fast decay as the light is switched off, and a slow component of the order of minutes. In the nuclear polarization EPR measurement (Fig.~3 of main text), this has the effect of modifying the relative polarization of each line as the experiment proceeds. Each point on the Fig.~\ref{fig:charge_transfer_correction} represents a full EPR spectrum, with the integrated intensities of two lines extracted by fitting. The data in the main text were corrected by normalizing to the total integrated area of each spectrum (and hence population of the defect center at that point in time).

\end{document}